\documentclass[11pt] {paper}
\usepackage{makeidx,graphics,graphicx,amssymb,amsfonts,epsfig}
\usepackage[english]{babel}
\usepackage{color}

\renewcommand{\vec}{\mathbf}

\newcommand{\ov}{\overline}

\begin{document} 

\title{A fuzzified BRAIN algorithm for learning DNF from incomplete data}


\author{
\renewcommand{\thefootnote}{\arabic{footnote}}
\rm Salvatore Rampone\footnotemark[1]  \and
\renewcommand{\thefootnote}{\arabic{footnote}}
\rm Ciro Russo\footnotemark[2]}
\maketitle
\footnotetext[1]{Dipartimento di Scienze per la Biologia, la Geologia e l'Ambiente. Universit\`{a} degli Studi del Sannio. Benevento, Italy. {\tt salvatore.rampone@unisannio.it} }
\footnotetext[2]{Dipartimento di Matematica. Universit\`{a} di Salerno. Italy. {\tt cirusso@unisa.it}}

\maketitle

\begin{abstract}
Aim of this paper is to address the problem of learning Boolean functions from training data with missing values. We present an extension of the BRAIN algorithm, called U-BRAIN (Uncertainty-managing Batch Relevance-based Artificial INtelligence), conceived for learning DNF Boolean formulas from partial truth tables, possibly with uncertain values or missing bits.

Such an algorithm is obtained from BRAIN by introducing fuzzy sets in order to manage uncertainty. In the case where no missing bits are present, the algorithm reduces to the original BRAIN.
\end{abstract}

\noindent {\bf Keywords}:
Boolean function,  DNF,  Learning algorithm,   Missing Values.


\section{Introduction}
\label{intro}
In many applications one must deal with data that have been collected incompletely. For example, in censuses and surveys, some participants may not respond to certain questions \cite{rubin1987}; in e-mail spam filtering, server information may be unavailable for e-mails from external sources \cite{Dicketal2008}; in medical studies, measurements on some subjects may be partially lost at certain stages of the treatment \cite{Ibrahim1990}; in DNA analysis, gene-expression microarrays may be
incomplete due to insufficient resolution, image corruption, or simply dust or scratches on the slide
\cite{Wangetal2006}; in sensing applications, a subset of sensors may be absent or fail to operate at
certain regions \cite{WilliamsandCarin2005}. 

Traditionally, data are often completed by ad hoc editing, such as case deletion and single imputation, where feature vectors with missing values are simply discarded or completed with specific values in the initial stage of analysis, before the main inference. Although analysis procedures designed for complete data become applicable after these edits, shortcomings are clear. For case deletion, discarding information is generally inefficient, especially when data are scarce. Secondly, the
remaining complete data may be statistically unrepresentative. More importantly, even if the incomplete-data problem is eliminated by ignoring data with missing features in the training phase, it is still inevitable in the test stage since test data cannot be ignored simply because a portion of features are missing. For single imputation, the main concern is that the uncertainty of the missing features is ignored by imputing fixed values \cite{wangetal2010}.

In this paper the general problem is posed as finding a Boolean function that
extends a partially defined one. This type of problem is studied, for example, in
learning theory \cite{pitt,val}, where it is called \emph{consistency problem}. In pattern recognition, a function separating two categories of data $T$ and $F$ is usually called a \emph{discriminant function} (e.g., \cite{manga}).
Apart from the existence of a solution for this problem --- which is a trivial question, if the self-consistency of the data is assumed --- it is desirable for such a solution to be presented in a canonical form (Boolean formulas in DNF or CNF, for example) and to be of minimum complexity \cite{blum}. It is well-known that such a problem, with a single positive given instance, is equivalent to a set covering problem \cite{ramp1,corm} and, therefore, NP-hard \cite{kear}. Nonetheless, it is possible to find a solution of approximately minimum complexity by applying generalized greedy methods (see, for example, \cite{john,haus}).


Here, as a further complication of the problem described above, we assume that training data might not be complete --- 
as previoulsly introduced
--- and the values of some elements of a given data vector may not be available. Usually a set of data which includes missing bits, is called a \emph{partially defined Boolean function with missing bits} \cite{bor}.

In order to deal with training data with incompletely observed attributes, we generalize the machine learning algorithm BRAIN (\emph{Batch Relevance-based Artificial INtelligence}) \cite{ramp1} for binary classification rules. This algorithm was originally conceived for recognizing \emph{splice junctions} in human DNA (see also \cite{ramp2,aloi}). Splice junctions are points on a DNA sequence at which ``superfluous'' DNA is removed during the process of protein synthesis in higher organisms \cite{green}. The general method used in the algorithm is related to the STAR technique of Michalski \cite{mic}, to the candidate-elimination method introduced by Mitchell \cite{mit}, and to the work of Haussler \cite{haus}.
Starting from the BRAIN algorithm, we extend it by using fuzzy sets \cite{zad}, in order to infer a DNF formula that is consistent with a given set of data which may have missing bits.

The paper is structured as follows. 

In Section \ref{sec:prob} we will formally describe the problem,  
recall some basics about BRAIN and  motivate the use of fuzzy sets.

In Section \ref{sec:alg} we shall describe the algorithm $\textrm{U-BRAIN}$ 
distinguishing --- for the sake of clearness --- three cases of increasing complexity. 

In Section \ref{sec:test} we shall apply the algorithm and discuss its performances on a standard dataset \cite{data}. We shall consider that missing bits in a dataset can appear for different reasons that can be essentially grouped in two cases. Some results are summarized in the tables contained in Appendix \ref{app}. 

Last, in Section \ref{sec:concl}, we will draw the conclusions.

\section{Problem description and formalization}
\label{sec:prob}

\subsection{Learning problem}
The problem of finding a Boolean function that extends a partially defined one can be posed as follows: given a pair of disjoint sets $$T, F \subseteq \{0,1\}^n$$ and a Boolean map $$f: G \to \{0,1\}$$ where $G = T \cup F$, such that $f[T] = \{1\}$ and $f[F] = \{0\}$, establish a Boolean function $$f^*: \{0,1\}^n \to \{0,1\}$$ such that $$f^*[T] = \{1\},  f^*[F] = \{0\}$$ 
The elements of $T$ and $F$ are usually called, respectively, \emph{positive} and \emph{negative instances}. 

\noindent
In \cite{ramp1} the author proposed an algorithm --- called BRAIN (Batch Relevance-based Artificial INtelligence) --- that infers a DNF Boolean formula in $n$ variables of low syntactic complexity and consistent with a given set of positive and negative instances. In other words, given a subset $G \subseteq \{0,1\}^n$ and a function $f: G \to \{0,1\}$, BRAIN yields a function $f^*: \{0,1\}^n \to \{0,1\}$, expressed with a DNF formula of approximately minimum complexity, such that $f^*$ coincide with $f$ on $G$.

Instances for which $f$ gives the value $1$ will be called \emph{positive}, those for which $f$ gives $0$ will be called \emph{negative}. In order to better distinguish positive and negative instances, we shall denote by $$\vec u_i, \quad i = 1, \ldots, p$$ the positive instances and by $$\vec v_j, \quad j = 1, \ldots, q$$ the negative ones.

In order to build a formula consistent with the given data, BRAIN compares each given positive instance $\vec u_i$ with each negative one $\vec v_j$ and builds a family of sets $S_{ij}$ --- that are (crisp) subsets of the set $L = \{x_i, \ov x_i \mid i = 1, \ldots, n\}$ of all literals --- as
$$S_{ij} = \left\{x_k \mid u_{ik} = 1, v_{jk} = 0\right\} \cup \left\{\ov x_k \mid u_{ik} = 0, v_{jk} = 1\right\},$$
or, that is the same, defined by their respective characteristic --- or membership --- functions
\begin{equation}\label{sijcrisp}
\begin{array}{l}
\chi_{ij}(x_k) = \left\{\begin{array}{ll} 1 & \textrm{if } u_{ik} = 1 \textrm{ and } v_{jk} = 0 \\ 0 & \textrm{otherwise} \end{array}\right. \\
\chi_{ij}(\ov x_k) = \left\{\begin{array}{ll} 1 & \textrm{if } u_{ik} = 0 \textrm{ and } v_{jk} = 1 \\ 0 & \textrm{otherwise} \end{array}\right. \\
\end{array}
\quad \forall k = 1, \ldots, n.
\end{equation}

Each $S_{ij}$ essentially represents a constraint that must be satisfied by the output formula; indeed it contains literals that are simultaneously positive in a positive instance and negative in a negative one. Therefore BRAIN, with a greedy approximation procedure, provides a formula that satisfies such conditions.

\subsection{Adding uncertainty}
Now we assume, in the same situation, that the given instances may contain ``uncertain'' values, i.e. for some elements of $G$ we may not know all the coordinates. This may occur as a consequence of errors or combination of data from different sources or, also, as a consequence of the application of a lossy compression algorithm.

Here we formalize such a cirumstance by representing $G$ as a subset of $\left\{0, \frac{1}{2}, 1\right\}^n$, where $\frac{1}{2}$ represents the uncertain values.
$$G \subseteq \left\{0, \frac{1}{2}, 1\right\}^n$$

Our aim, now, is to extend BRAIN in such a way that the new algorithm would 
infer a function $f^*: \{0,1\}^n \to \{0,1\}$, expressed by a DNF formula of minumum complexity, that is consistent with $f$ in the following sense:
\begin{itemize}
\item $f^*$ coincide with $f$ on $G \cap \{0,1\}^n$,
\item for any vector $\vec u = (u_1, \ldots, u_n) \in G \cap \left\{0, \frac{1}{2}, 1\right\}^n$, there exists an element $\vec u' = (u'_1, \ldots, u'_n) \in \{0,1\}^n$ such that $u_i = u'_i$ for any $u_i \neq \frac{1}{2}$ and $f^*(\vec u') = f(\vec u)$.
\end{itemize}

In what follows, by an \emph{instance} we shall mean an element of $\left\{0, \frac{1}{2}, 1\right\}^n$ and, if it is in $\{0,1\}^n$ it will be called \emph{certain}. 

As we shall see, in our extensions of BRAIN, the sets $S_{ij}$ shall be fuzzy subsets of $L$.

In writing DNF formulas, throughout the paper we shall often omit the conjunction symbol $\wedge$ and we shall denote by an overlined letter the negation of a variable, in order to make formulas more compact and readable; thus we may write, for instance, $x_1 \ov x_3 \vee \ov x_2 x_4$ instead of $(x_1 \wedge \lnot x_3) \vee (\lnot x_2 \wedge x_4)$.

\section{Algorithm}
\label{sec:alg}

\subsection{Uncertainty reduction}
\label{subsec:uncred}

The first step of our algorithms aims at reducing as much as possible the number of missing bits. Indeed we may have instances in which missing bits can be recovered. Assume, for example, to have a positive instance with a single missing bit and a negative one with no missing bits that coincide on all of their certain values. In this case, since we assume the given instances to be self-consistent, the missing bit in the positive instance must give the only possible difference between the two instances, whence it must be the negation of the corresponding bit in the negative instance.

Therefore, as a first step that is common to both the algorithms, we shall update the data as follows.

Let $(\vec u,\vec v)$ be a pair of instances, one positive and the other one negative, and assume that there exists a unique coordinate $k \in \{1, \ldots, n\}$ such that
\begin{itemize}
\item $u_r = v_r \in \{0,1\}$ for all $r \neq k$,
\item one of the two $k$-th components $u_k, v_k$ is certain and the other one is not.
\end{itemize}
In this case we must have $u_k = \ov v_k$, hence we update the instance containing an uncertainty by substituting its $k$-th component with the negation of the $k$-th component of the certain instance.

This substitution shall be made whenever possible and the reduction iterated until no more substitutions of this kind are possible.

\noindent {\bf Example}
Let $G = \{\vec u_1, \vec v_1, \vec v_2\}$, with
$$\vec u_1 = (1, 1/2, 0, 0), \vec v_1 = (1,1,0,0), \vec v_2 = (1,0,0,1/2).$$
>From the comparison of $\vec u_1$ and $\vec v_1$ we obtain that the only Boolean value for $u_{12}$ that keeps the set of instances self-consistent is $\ov v_{12}$, i.e. 0; so we set $\vec u'_1 = (1,0,0,0)$.

Once we update the set of instances by substituting $\vec u_1$ with $\vec u'_1$, we can apply the same argument to $u'_1$ and $\vec v_2$ thus obtaining $v'_{24} = 1$ and $\vec v'_2 = (1,0,0,1)$. Therefore the new set of instances is $G' = \{\vec u'_1, \vec v'_1, \vec v'_2\}$.

After the description of the algorithms we will show, in Example \ref{rednec}, the importance of the reduction.

\subsection{Repetition deletion}
\label{subsec:repdel}

It is possible that the set of instances contains redundant information, i.e. there are some 
instances that are repeated one or more times, either since the beginning or as a result of the reduction step.

Such redundancy is removed by keeping each certain instance just once and deleting all the repetitions.

\subsection{Membership function}
\label{subsec:builfor}


We will consider the fuzzy subsets $S_{ij}$ 
of $L$ defined by the 
characteristic function $\chi_{ij}$ 
:
\begin{equation}\label{chi2}
\begin{array}{l}
\begin{array}{l}
\chi_{ij}(x_k) = 
	\left\{\begin{array}{ll} 
			1 & \textrm{if } u_{ik} = 1 \textrm{ and } v_{jk} = 0 \\
			\left(\frac{1}{2}\right)^{p+q} & \textrm{if } u_{ik} > v_{jk} \textrm{ and } \frac{1}{2} \in \{u_{ik}, v_{jk}\} \\
			\left(\frac{1}{2}\right)^{p+q+1} & \textrm{if } u_{ik} = v_{jk} = \frac{1}{2} \\
			0 & \textrm{otherwise}
		\end{array}\right. \\
\chi_{ij}(\ov x_k) =
	\left\{\begin{array}{ll} 
			1 & \textrm{if } u_{ik} = 0 \textrm{ and } v_{jk} = 1 \\
			\left(\frac{1}{2}\right)^{p+q} & \textrm{if } u_{ik} < v_{jk} \textrm{ and } \frac{1}{2} \in \{u_{ik}, v_{jk}\} \\
			\left(\frac{1}{2}\right)^{p+q+1} & \textrm{if } u_{ik} = v_{jk} = \frac{1}{2} \\
			0 & \textrm{otherwise}
		\end{array}\right. \\
\end{array} \\ \\
\forall k = 1, \ldots, n.
\end{array}
\end{equation}

Such a membership function is thought of in such a way that the certain values
have a much prominent role w.r.t. the missing ones in a comparison between two
instances. However, as we shall see later on, even uncertain information may gain
relevance if the certain one is not sufficient.

\subsection{One-to-one}
\label{subsubsec:1-1}
For the sake of clearness, we shall distinguish three situations, where we are given, respectively
\begin{itemize}
\item a positive instance and a negative one,
\item a positive instance and several negative ones,
\item several positive and negative instances.
\end{itemize}

First let us find a DNF formula that is consistent with a positive instance $\vec u_i$ and a negative one $\vec v_j$. A term $t$ made true by $\vec u_i$ and false by $\vec v_j$ must include at least one variable not assigned in the same way in the two instances. 
By the assumption of self-consistency of our instances, $\chi_{ij}$ is not constantly equal to 0. Now, if $l$ is a literal such that $\chi_{ij}(l) = \max \chi_{ij}$, $f^* = l$ is consistent with the given set of instances and of minimum complexity. If there is more than one literal satisfying such a condition, we choose the positive one with the lowest index --- if any --- or, otherwise, the negative one with the lowest index. By the membership function definition (2), the certain differences are necessarily privileged.

\noindent {\bf Example}\label{eq11}
Let
$$\vec u_1 = (1,1,0,1/2,1), \quad \vec v_1 = (1,0,0,1,0).$$
Then
$$\chi_{11}(l) = \left\{\begin{array}{ll}
 1 & \textrm{for } l \in \{x_2, x_5\} \\
 1/4 & \textrm{for } l = \ov x_4 \\
 0 & \textrm{otherwise}
 \end{array}\right..$$
Therefore $f^* = x_2$.  

\noindent {\bf Example}\label{diff11}
Let $\vec u_1 = (1,1/2,0)$ and $\vec v_1 = (1/2,0,0)$. Then
$$\chi_{11}(l) = \left\{\begin{array}{ll}
 1/4 & \textrm{for } l \in \{x_1, x_2\} \\
 0 & \textrm{otherwise}
 \end{array}\right..$$
Therefore 
$f^* = x_1$.

It is worth noting the $x_1$ selection in the last example implies a decision on the negative instance, that is implicitly assumed to be $v^*_1 = (1,0,0)$.

\subsection{One-to-many}
\label{subsubsec:1-q}

Now assume we have a positive instance $\vec u_i$ and $q$ negative ones $\vec v_1, \ldots, \vec v_q$. In this case, the problem is equivalent to a set covering one and we follow the BRAIN algorithm or, in a different version, the set covering approximation improvement presented in \cite{aloi}. Recalling that, for a fuzzy subset $F$ of a given set $S$, the \emph{fuzzy cardinality} is defined as
\begin{equation}\label{fcard}
\sharp(F) = \displaystyle\sum_{x \in S} \chi_F(x),
\end{equation}
we define the \emph{relevance} of a literal $l_k$ in $S_{ij}$ as
\begin{equation}\label{rij}
R_{ij}(l_k) = \frac{\chi_{ij}(l_k)}{\sharp(S_{ij})},
\end{equation}
where $\chi_{ij}$ is the membership function of $S_{ij}$. Then we set
\begin{equation}\label{ri}
R_i(l_k) = \frac{1}{q}\sum_{j=1}^q R_{ij}(l_k).
\end{equation}
So, given the $S_{ij}$ sets, we proceed as follows.
\begin{enumerate}
\item Compute the relevances of each literal (the relevance of the opposite of the previously selected literals is always set to zero). 
\item Choose the literal $l$ with the highest relevance.
\item Erase the sets containing $l$ and the occurrences of $\ov l$ inside the sets where they appear.
\item Repeat steps 1--3 until there are no more sets.
\end{enumerate}

The resulting formula will be the conjunction of the chosen literals  $l_{i_1}, \ldots, l_{i_k}$ and is obviously consistent with $\vec u_1$. On the other hand, for each $j \leq q$, $\vec v_j$ either contains a certain value which is in contradiction with one of such literals, or contains an uncertain value that is assumed, by the algorithm, contradicting a literal of $f^*$. Then the $\vec v_j$'s are indeed negative instances of $f^*$.

\noindent {\bf Example}\label{eq1m}
Let $\vec u_1 = (1,0,0)$, $\vec v_1 = (0,1,1)$, $\vec v_2 = (1,0,1)$ and $\vec v_3 = (1,1/2,1)$. The $S_{1j}$ are crisp sets:
$$S_{11} = \{x_1, \ov x_2, \ov x_3\}, \quad S_{12} = \{\ov x_3\}, \quad S_{13} = \{\ov x_2, \ov x_3\}.$$
%
where
$$\chi_{11}(l) = \left\{\begin{array}{ll}
 1 & \textrm{for } l = x_1, \ov x_2, \ov x_3 \\
 0 & \textrm{otherwise}
 \end{array}\right.$$

$$\chi_{12}(l) = \left\{\begin{array}{ll}
 1 & \textrm{for } l = \ov x_3 \\
 0 & \textrm{otherwise}
 \end{array}\right.$$
and
$$\chi_{13}(l) = \left\{\begin{array}{ll}
 1 & \textrm{for } l = \ov x_3 \\
 1/32 & \textrm{for } l = \ov x_2 \\
 0 & \textrm{otherwise}
 \end{array}\right..$$
The maximum resulting relevance is $R_1(\ov x_3)$ and $\chi_{1j}(\ov x_3) > 0$ for all $j$; it follows $f^* = \ov x_3$.

\noindent {\bf Example}\label{diff1m}
Let
$$\vec u_1 = (1,0,1/2,1), \quad \vec v_1 = (0,1,1,1), \quad \vec v_2 = (1,0,1,0).$$
The $S_{1j}$'s are, again, crisp sets:
$$S_{11} = \{x_1, \ov x_2, \ov x_3\}, \quad S_{12} = \{x_4, \ov x_3\}.$$

$$\chi_{11}(l) = \left\{\begin{array}{ll}
 1 & \textrm{for } l \in \{x_1, \ov x_2\} \\
 1/8 & \textrm{for } l = \ov x_3 \\
 0 & \textrm{otherwise}
 \end{array}\right.$$
and
 $$\chi_{12}(l) = \left\{\begin{array}{ll}
 1 & \textrm{for } l = x_4 \\
 1/8 & \textrm{for } l = \ov x_3 \\
 0 & \textrm{otherwise}
 \end{array}\right.,$$
hence $\max R_1 = R_1(x_4)$. Once we choose $x_4$, we erase $S_{12}$, that is the only set in which the membership function of $x_4$ is not 0. Then we choose $x_1$ in $S_{11}$ and the resulting formula is $f^* = x_4 \wedge x_1$.

\subsection{Many-to-Many}
\label{subsubsec:p-q}

Let us now consider the case of $p$ positive instances and $q$ negative ones:
$$\vec u_1, \ldots, \vec u_p, \vec v_1, \ldots, \vec v_q.$$
A consistent DNF formula will be a disjunction of a set of conjunction (or product) terms, i.e. of a set of conjunctions of literals. Each positive instance shall satisfy at least one product term and none of the terms shall be satisfied by any of the negative instances. So, for each pair $(\vec u_i,\vec v_j)$, there exists a term that is satisfied by $\vec u_i$ and, like all the other terms, is not satisfied by $\vec v_j$; this implies that such a term must contain a variable that is not assigned in the same way in $\vec u_i$ and $\vec v_j$, which means that it contains a literal of $S_{ij}$. Then, for all $i = 1, \ldots, p$ and $j = 1, \ldots, q$, there  exists a term in $f^*$ containing a literal in $S_{ij}$.

In this case we have the $pq$ sets $S_{ij}$ that we collect, for our convenience, as
$$S_i = \{S_{i1}, \ldots, S_{iq}\}, \qquad i = 1, \ldots, p,$$
with the associated relevance functions $R_i$ defined by (\ref{ri}) and the total relevance defined by
\begin{equation}\label{r}
R(l_k) = \frac{1}{p} \sum_{j = 1}^p R_i(l_k) = \frac{1}{pq} \sum_{j = 1}^p \sum_{i=1}^q R_{ij}(l_k).
\end{equation}
The U-BRAIN algorithm uses the relevance as a greedy criterion to build DNF function terms. Starting from an empty term, it selects the most relevant variable and add it to the term, erasing all the satisfied constraints $S_{ij}$ and all the incompatible $S_i$ sets, until at least a set $S_i$ of constraints is satisfied (empty). This greedy choice aims at the same time both at covering the greatest number of positive instances and at selecting the less possible number of variables. The process is iterated until there are no more constraints.

\subsection{Negative Instances Updating}
Each time a term is produced, the implicit choices over the uncertain components of the negative instances, if any, must be explicited to avoid contradiction with the terms to be generated in the following. So for each negative instance $\vec v_j$, if a particular choice of the uncertain values can satisfy the last generated term $m$, 
i.e. there exists an element $$\vec v' = (v'_1, \ldots, v'_n) \in \{0,1\}^n$$ such that $$v_{ij} = v'_i$$ for any $v_{ij} \neq \frac{1}{2}$ and $$m(\vec v') = 1$$
then the uncertain element of lowest index of $\vec v_j$ is set to a certain value contradicting the truthness of the term. Once the negative instances have been updated, the
algorithm checks the self-consistency of the new set of instances and, in case a consistency issue arises, it stops.

The whole U-BRAIN algorithm can be formally depicted as in figure 1.

\begin{figure}

\begin{enumerate}
\item[]$\textbf{U-BRAIN algorithm}$
\item[{\bf 1}] \emph{Input:}
	\begin{itemize}
	\item $n$ the number of variables,
	\item $G = \{\vec u_1, \vec u_2, \ldots , \vec u_p, \vec v_1, \vec v_2, \ldots, \vec v_q\}$ the set of training instances.
	\end{itemize}
		\emph{Initialization:} set $f^* = \phi$.
\item[{\bf 2}] While there are positive instances in $G$
	\begin{enumerate}
	\item[2.0] Uncertainty reduction, repetition deletion.
	\item[2.1] \emph{$S_{ij}$ sets:} Build from $G$ the $S_{ij}$ sets and collect them in $S_i = \{S_{ij}\}_{j=1}^q$ for all $i = 1, \ldots, p$.
	\item[2.2] \emph{Start a new term:} Set $m = \phi$.
	\item[2.3] \emph{Build the term:} While there are $S_{ij}$ sets
		\begin{enumerate}
		\item[{\em 2.3.1}] \emph{Relevances:} Compute the relevance $R(l_k)$, for $l_k \in \{x_k, \ov x_k\}$, $k = 1, \ldots, n$;
		\item[{\em 2.3.2}] \emph{Add variable:} Select $l_k$ such that $R(l_k)$ is equal to $\max R$, $m \leftarrow m \wedge l_k$;
		\item[{\em 2.3.3}] \emph{Update sets:} Erase the $\ov l_k$ occurrences, if any, in the $S_i$ sets where also $l_k$ appears, erase the $S_i$ sets in which $l_k$ does not occur, erase the $S_{ij}$ sets where $l_k$ appears.\footnote{By a literal that \emph{appears} or \emph{occurs} in an $S_{ij}$ we mean that its value under the $\textit{ij}$-th membership function is $> 0$; a literal appears in $S_i$ if it appears in $S_{ij}$ for some $j$.}
		\end{enumerate}
	\item[2.4] \emph{Add the term:} $f^* \leftarrow f^* \vee m$.
	\item[2.5] \emph{Update positive instances:} Erase from $G$ all the positive instances satisfying $m$. 
	\item[2.6] \emph{Update negative instances:} Update uncertain values $\forall \vec v_j \in G$. 
	\end{enumerate}
\item[{\bf 3}] \emph{Output:} $f^*$. 
\end{enumerate}

\caption[]{U-BRAIN algorithm.}
\end{figure}

It is worth noticing that, if the given set of instances is contained in $\{0,1\}^n$, then the algorithm proposed coincides with the original BRAIN.

As we anticipated in subsection \ref{subsec:uncred}, the following example shows the importance of the reduction step.

\noindent {\bf Example}\label{rednec}
Let $\vec u_1 = (1,0,0)$, $\vec u_2 = (1/2,1,0)$ be positive instances and $\vec v_1 = (1,1/2,0)$ a negative instance.

If we apply the reduction, first $\vec v_1$ shall be substituted by $\vec v'_1 = (1,1,0)$ and, then, $\vec u_2$ will be substituted by $\vec u'_2 = (0,1,0)$. Then the algorithm will give us the formula $\ov x_1 \vee \ov x_2$ that is consistent with the given set of instances.

If we do not apply the reduction, both the algorithms will give back the formula $\ov x_2 \vee x_2$ which is a tautology and, therefore, is not consistent with our negative instance.

\section{Performance evaluations}
\label{sec:test}

We shall distinguish two cases, that essentially depend on the reason of uncertainty. More precisely, we may have uncertain values due to the following circumstances:
\begin{enumerate}
\item \emph{random uncertainties}, due e.g. to the presence of some noise or errors in data storage, transmission and/or retrieving,
\item \emph{trustworthy uncertainties}, occurring, for instance, when missing bits have not been provided because some of the sources recognize them as not relevant.
\end{enumerate}

These two aspects of incomplete information appear to naturally correspond to
fuzzy logic and probability respectively \cite{dp}.

Now, since our aim is to treat both these situations, the best approach seems
to require both fuzzy logic and probability in some sense. So, in the definition of the $\chi_{ij}$ sets, we privileged the logical viewpoint assuming that the missing bit is a vague information that must be used only when no better one is present. On the other hand, the possibility that missing bits come from randomly distributed errors
is taken care of at a subsequent stage, namely when we define the relevance of a literal (subsections \ref{subsubsec:1-q} and \ref{subsubsec:p-q}). In fact the total relevance defined by (\ref{r}) is indeed
a probability distribution and its definition by means of the fuzzy cardinality (\ref{fcard}) gives some weight back to missing bits when necessary.

So the algorithms have been tested using the standard ``Zoo'' dataset in \cite{data}, with the introduction of random and trustworthy missing bits alternatively.

The dataset contains one hundred one animals divided in seven types (mammal, bird, reptile, fish, amphibian, insect, invertebrate) and described by means of fifteen Boolean attributes (hair, feathers, eggs, milk, airborne, aquatic, predator, toothed, backbone, breathes, venomous, fins, tail, domestic, catsize) and a numeric one (legs = 0,2,4,5,6, or 8) that we made into five further Boolean values (legs = 2, legs = 4, legs = 5, legs = 6, legs = 8), with legs = 0 represented by assigning the value 0 to all of them.

The tests have been performed by considering the animals of a given type as positive instances and all the others as negative ones, and then computing $f^*$ by means of U-BRAIN. Eventually, missing bits have been randomly distributed inside the data, up to 50\% of the total number of bits; analogously, the trustworthy uncertainties have been distributed among the data, up to 50\% of the total number of bits.

We use two indicators of the performances. The \emph{Average Error Number} $AEN$ is the average number of instances in the dataset erroneously classified. The \emph{error rate} $R$ is the ratio between the error number and the dataset size. 

In Table \ref{tab} such indicators are reported while, in Appendix (Tables \ref{type1}--\ref{type7}) we present some results of such performances; each ``E'' column indicates the number of errors introduced by the algorithm. More precisely, it shows the number of instances that the resulting formula changes from positive to negative or viceversa.

\begin{table}[!ht]
\caption{Overall Average Error Number and Error rate (Zoo Dataset) varying the missing bit percentage in the Random and Trustworthy cases.}
\centering
\begin{tabular}{|c|c|c|}
\hline
 Random& $AEN$ & $R$ \\ \hline
 10\%  & $0.36$ & $0.003$    \\ \hline
 20\%  	& $0.93$ 	& $0.009$   \\ \hline
 30\% 	& $1.00$ 	& $0.010$    \\ \hline
 40\%  	& $1.50$ 	& $0.015$    \\ \hline
 50\%  	& $5.43$ 	& $0.054$    \\ \hline
\multicolumn{3}{|c|}{} \\ \hline
Trustworthy			 & $AEN$ & $R$  \\ \hline
 10\%  	& $0.57$ 	& $0.006$    \\ \hline
 20\%  	& $0.28$ 	& $0.003$   \\ \hline
 30\%  	& $0.28$ 	& $0.003$    \\ \hline
 40\%  	& $0$ 	& $0$    \\ \hline
 50\%  	& $0$ 	& $0$    \\ \hline
\end{tabular}\label{tab}
\end{table}

In the case of random missing bits, the average error percentage is below $1/10$ of the percentage of missing bits, showing a highly reliable behaviour. 
To what extent the case of trustworthy uncertainties, as Table \ref{tab} shows, the results obtained by $\textrm{U-BRAIN}$ show the number of errors decreases as the number of missing bits increases. This circumstance suggests the possibility of applying $\textrm{U-BRAIN}$ for the reconstruction of highly compressed data, which is a motivation for future works.

\section{Concluding remarks}
\label{sec:concl}

The problem addressed in this paper is to find, given a partially defined Boolean function with missing bits, a Boolean formula in disjuctive normal form, of approximately minimum complexity, that is consistent with the given data.

The solutions proposed is a learning algorithm --- obtained as extension of the BRAIN algorithm by means of the introduction of fuzzy sets --- inferring Boolean formulas from incomplete instances. The conjunctive terms of the formula are computed in an iterative way by identifying, from the given data, a family of sets of conditions that must be satisfied by all the positive instances and violated by all the negative ones; such conditions allow the computation of a relevance coefficient for each attribute (literal). 

The proposed approach introduces the possibility of managing uncertain values by considering the aforementioned sets of conditions as fuzzy sets, whose characteristic functions play a significant role in the relevance coefficients.

The new algorithms appear to have low error rates and maintain the polynomial computational complexity of the original BRAIN algorithm.

\newpage

\appendix

\section{Test tables}
\label{app}

\begin{table}[htp]
\caption{Type 1 -- Mammal: $f = x_4$}
\centering
\begin{tabular}{|c|c|c|}
\hline
\multicolumn{3}{|c|}{Random (1)} \\ \hline
  & $f^*$  & E \\ \hline
 10\%  & $x_4$ & 0\\ \hline
 20\%  & $x_4$ & 0\\ \hline
 30\%  & $x_4$ & 0\\ \hline
 40\%  & $x_4$ & 0\\ \hline
 50\% &	$\ov x_3 \ov x_{11} \vee x_{16} x_{19}$	& 1\\ \hline
\multicolumn{3}{|c|}{Random (2)} \\ \hline
   & $f^*$  & E \\ \hline
 10\%  & $x_4$ & 0\\ \hline
 20\%  & $x_4$ & 0\\ \hline
 30\%  & $x_4$ & 0\\ \hline
 40\%  & $x_4$ & 0\\ \hline
 50\%  & $x_1 x_8 \vee \ov x_3 x_8$	& 1\\ \hline
\multicolumn{3}{|c|}{Trustworthy} \\ \hline
   & $f^*$  & E \\ \hline
 10\%  & $x_4$ & 0\\ \hline
 20\%  & $x_4$ & 0\\ \hline
 30\%  & $x_4$ & 0\\ \hline
 40\%  & $x_4$ & 0\\ \hline
 50\%  & $x_4$ & 0\\ \hline
\end{tabular}\label{type1}
\end{table}

\begin{table}[htp]
\caption{Type 2 -- Bird: $f = x_2$}
\centering
\begin{tabular}{|c|c|c|}
\hline
\multicolumn{3}{|c|}{Random (1)} \\ \hline
   & $f^*$  & E \\ \hline
 10\%  & $x_2$ & 0 \\ \hline
 20\%  & $x_2$ & 0 \\ \hline
 30\%  & $x_2$ & 0 \\ \hline
 40\%  & $x_{17} \ov x_8$ & 0 \\ \hline
 50\% & $x_{17} \ov x_4$ & 0 \\ \hline
\multicolumn{3}{|c|}{Random (2)} \\ \hline
   & $f^*$  & E \\ \hline
 10\%  & $x_2$ & 0 \\ \hline
 20\%  & $x_2$ & 0 \\ \hline
 30\%  & $x_2$ & 0 \\ \hline
 40\%  & $x_{17} \ov x_1$ & 0 \\ \hline
 50\%  & $x_2$ & 0 \\ \hline
\multicolumn{3}{|c|}{Trustworthy} \\ \hline
   & $f^*$  & E \\ \hline
 10\%  & $x_2$ & 0 \\ \hline
 20\%  & $x_2$ & 0 \\ \hline
 30\%  & $x_2$ & 0 \\ \hline
 40\%  & $x_2$ & 0 \\ \hline
 50\%  & $x_2$ & 0 \\ \hline
\end{tabular}\label{type2}
\end{table}

\begin{table}[htp]
\caption{Type 3 -- Reptile: $f = \ov x_6 \ov x_1 x_8 \vee x_{11} \ov x_3 x_6 \vee x_{16} \ov x_8 \ov x_6$}
\centering
\begin{tabular}{|c|c|c|}
\hline
\multicolumn{3}{|c|}{Random (1)} \\ \hline
  & $f^*$  & E \\ \hline
 10\% & $\ov x_1 x_8 \ov x_{12} x_{18} \ov x_6 $ &  \\ 
 & $\vee x_{11} x_8 x_{18} \vee x_{16} \ov x_8 \ov x_1$  & 2 \\ \hline
 20\%  & $x_{11} x_8 \ov x_{16} \vee x_{16} \ov x_1 x_{20}$ & 4 \\ \hline
 30\%  & $\ov x_1 x_8 \ov x_{12} \vee x_{20} \ov x_8 \ov x_7$ & 7 \\ \hline
 40\%  & $\ov x_1 x_8 x_{11} \vee x_{16} x_3 \ov x_6$ & 3 \\ \hline
 50\% & $\ov x_4 x_8 x_7 \vee \ov x_8 x_{20}$ & 20 \\ \hline
\multicolumn{3}{|c|}{Random (2)} \\ \hline
 & $f^*$  & E \\ \hline
 10\%  & $x_{11} x_{18} \ov x_{20} \vee x_{16} \ov x_1 \ov x_6$ & 2 \\ \hline
 20\% & $\ov x_1 x_8 \ov x_{12} x_7 \vee x_{16} \ov x_8 x_9$ & 4 \\ \hline
 30\% & $\ov x_1 \ov x_6 x_9 \ov x_2 \vee x_{11} x_{18} \ov x_{12}$ & 1 \\ \hline
 40\%  & $\ov x_1 x_{16} \vee \ov x_4 x_8 x_{17}$ & 8 \\ \hline
 50\% & $\ov x_5 x_3 x_{10} \vee x_{11} \ov x_3$ &  14 \\ \hline
\multicolumn{3}{|c|}{Trustworthy} \\ \hline
   & $f^*$  & E \\ \hline
 10\%  & $x_{15} x_8 \vee x_{16} \ov x_8 \ov x_6$ & 4 \\ \hline
 20\% & $\ov x_1 \ov x_6 x_8 \vee \ov x_1 x_{15} x_{16}$ & \\ 
 & $ \vee x_{11} \ov x_3 x_6$ & 1 \\ \hline
 30\% & $\ov x_1 \ov x_6 x_8 \vee \ov x_1 x_{15} x_{16}$ & \\
 & $ \vee x_{11} \ov x_3 x_6$ & 1 \\ \hline
 40\% & $\ov x_6 \ov x_1 x_8 \vee x_{11} \ov x_3 x_6$ & \\ 
 & $ \vee x_{16} \ov x_8 \ov x_6$ & 0 \\ \hline
 50\% & $\ov x_1 \ov x_6 x_8 \vee x_{11} \ov x_3 x_6$ & \\ 
 & $ \vee x_{16} \ov x_8 \ov x_6$ & 0 \\ \hline
\end{tabular}\label{type3}
\end{table}

\begin{table}[htp]
\caption{Type 4 -- Fish: $f = x_{12} x_3$}
\centering
\begin{tabular}{|c|c|c|}
\hline
\multicolumn{3}{|c|}{Random (1)} \\ \hline
 & $f^*$  & E \\ \hline
 10\%  & $\ov x_{10} x_8 x_3$ & 0 \\ \hline
 20\%  & $x_{12} x_3$ & 0 \\ \hline
 30\%  & $\ov x_{10} x_{18}$ & 1 \\ \hline
 40\%  & $\ov x_{10} x_{18}$ & 1 \\ \hline
 50\%  & $\ov x_{10} x_9$ & 1 \\ \hline
\multicolumn{3}{|c|}{Random (2)} \\ \hline
 & $f^*$  & E \\ \hline
 10\%  & $x_{12} x_3$ & 0 \\ \hline
 20\%  & $\ov x_{10} x_{12}$ & 0 \\ \hline
 30\%  & $x_{12} \ov x_4$ & 0 \\ \hline
 40\%  & $x_{12} \ov x_4$ & 0 \\ \hline
 50\% & $x_{12} \ov x_{20} \vee x_{12} \ov x_5$ & 5 \\ \hline
\multicolumn{3}{|c|}{Trustworthy} \\ \hline
  & $f^*$  & E \\ \hline
 10\%  & $x_{12} x_3$ & 0 \\ \hline
 20\% & $x_{12} x_3$ & 0 \\ \hline
 30\%  & $x_{12} x_3$ & 0 \\ \hline
 40\%  & $x_{12} x_3$ & 0 \\ \hline
 50\%  & $x_{12} x_3$ & 0 \\ \hline
\end{tabular}\label{type4}
\end{table}

\begin{table}[htp]
\caption{Type 5 -- Amphibian: $f = x_{16} x_6 x_8 x_3$}
\centering
\begin{tabular}{|c|c|c|}
\hline
\multicolumn{3}{|c|}{Random (1)} \\ \hline
 &$f^*$  & E \\ \hline
 10\%  & $x_{16} x_6 x_8$ & 1 \\ \hline
 20\%  & $x_6 x_{10} \ov x_4 \ov x_{17}$ & 0 \\ \hline
 30\%  & $x_6 x_{16} x_9$ & 2 \\ \hline
 40\%  & $x_6 x_{16} x_8$ & 1 \\ \hline
 50\%  & $\ov x_{18} x_9$ & 7 \\ \hline
\multicolumn{3}{|c|}{Random (2)} \\ \hline
   & $f^*$  & E \\ \hline
 10\%  & $x_{16} x_6 \ov x_{20}$ & 1 \\ \hline
 20\%  & $\ov x_{18} x_9 x_3 \vee x_6 x_{16} x_{10}$ & 2 \\ \hline
 30\%  & $x_{16} \ov x_1 x_8$ & 1 \\ \hline
 40\%  & $x_{16} x_3 x_8$ & 1 \\ \hline
 50\%  & $x_8 \ov x_1 x_{18} \vee x_{16} \ov x_1 x_8$ & 19 \\ \hline
\multicolumn{3}{|c|}{Trustworthy} \\ \hline
   & $f^*$  & E \\ \hline
 10\%  & $x_6 x_{16} x_8 x_3$ & 0 \\ \hline
 20\%  & $x_{16} x_6 x_8 x_3$ & 0 \\ \hline
 30\%  & $x_{16} x_6 x_8 x_3$ & 0 \\ \hline
 40\%  & $x_6 x_{16} x_8 x_3$ & 0 \\ \hline
 50\%  & $x_{16} x_6 x_8 x_3$ & 0 \\ \hline
\end{tabular}\label{type5}
\end{table}

\begin{table}[htp]
\caption{Type 6 -- Insect: $f = x_{14} x_{10}$}
\centering
\begin{tabular}{|c|c|c|}
\hline
\multicolumn{3}{|c|}{Random (1)} \\ \hline
 & $f^*$  & E \\ \hline
 10\%  & $x_{14} \ov x_6$ & 0 \\ \hline
 20\%  & $x_{14} \ov x_7$ & 1 \\ \hline
 30\%  & $x_{14} x_{10}$ & 0 \\ \hline
 40\%  & $\ov x_8 \ov x_2 \ov x_7 \vee x_5 \ov x_2 \ov x_9$ & 2 \\ \hline
 50\%  & $\ov x_9 x_{10} x_{14}$ & 0 \\ \hline
\multicolumn{3}{|c|}{Random (2)} \\ \hline
   & $f^*$  & E \\ \hline
 10\%  & $x_{14} \ov x_6$ & 0 \\ \hline
 20\%  & $\ov x_9 x_{14} x_{10}$ & 0 \\ \hline
 30\%  & $x_{14} x_{10}$ & 0 \\ \hline
 40\%  & $\ov x_9 \ov x_7$ & 4 \\ \hline
 50\%  & $x_{14} \ov x_7$ & 1 \\ \hline
\multicolumn{3}{|c|}{Trustworthy} \\ \hline
   & $f^*$  & E \\ \hline
 10\%  & $x_{14} x_{10}$ & 0 \\ \hline
 20\%  & $x_{14} x_{10}$ & 0 \\ \hline
 30\%  & $x_{14} x_{10}$ & 0 \\ \hline
 40\%  & $x_{14} x_{10}$ & 0 \\ \hline
 50\%  & $x_{14} x_{10}$ & 0 \\ \hline
\end{tabular}\label{type6}
\end{table}

\begin{table}[htp]
\caption{Type 7 -- Invertebrate: $f = \ov x_9 \ov x_{14} \vee \ov x_{10} x_{14}$}
\centering
\begin{tabular}{|c|c|c|}
\hline
\multicolumn{3}{|c|}{Random (1)} \\ \hline
   & $f^*$  & E \\ \hline
 10\%  & $\ov x_9 \ov x_{10} \vee \ov x_9 \ov x_{14}$ & 0 \\ \hline
 20\%  & $\ov x_9 \ov x_{14} \vee x_{14} \ov x_{10}$ & 0 \\ \hline
 30\% & $\ov x_9 x_6 \vee \ov x_9 \ov x_1 x_7 \vee \ov x_9 \ov x_1 \ov x_5$ & 2 \\ \hline
 40\%  & $\ov x_9 \ov x_{10} \vee \ov x_9 \ov x_{14}$ & 0 \\ \hline
 50\%  & $\ov x_9 x_7$ & 3 \\ \hline
\multicolumn{3}{|c|}{Random (2)} \\ \hline
   & $f^*$  & E \\ \hline
 10\%  & $\ov x_9 \ov x_{14} \vee \ov x_9 x_6$ & 0 \\ \hline
 20\%  & $\ov x_9 \ov x_5$ & 2 \\ \hline
 30\%  & $\ov x_9 \ov x_{14} \vee x_{14} \ov x_{10}$ & 0 \\ \hline
 40\%  & $\ov x_9 \ov x_{14} \vee x_{14} x_7$ & 1 \\ \hline
 50\%  & $\ov x_9 \ov x_1$ & 4 \\ \hline
\multicolumn{3}{|c|}{Trustworthy} \\ \hline
   & $f^*$  & E \\ \hline
 10\%  & $\ov x_9 \ov x_{14} \vee \ov x_{10} x_{14}$ & 0 \\ \hline
 20\%  & $\ov x_9 \ov x_{14} \vee x_{14} x_6$ & 1 \\ \hline
 30\%  & $\ov x_9 \ov x_{14} \vee x_{14} \ov x_{10}$ & 0 \\ \hline
 40\%  & $\ov x_9 \ov x_{14} \vee x_{14} \ov x_{10}$ & 0 \\ \hline
 50\%  & $\ov x_9 \ov x_{14} \vee x_{14} \ov x_{10}$ & 0 \\ \hline
\end{tabular}\label{type7}
\end{table}


\begin{thebibliography}{99}

\bibitem{aloi}
Aloisio A., Izzo V., Rampone S., VLSI implementation of greedy-based distributed routing schemes for ad hoc networks, {\em Soft Computing}, {\bf 11}/9, 865--872, 2007.

\bibitem{blum}
Blumer A., Ehrenfeucht A., Haussler D., Warmuth M.K., Occam's Razor, {\em Information Processing Letters}, {\bf 24}, 377--380, 1987.

\bibitem{bor}
Boros E., Ibaraki T., Makino K., Logical analysis of binary data with missing bits, {\em Artificial Intelligence}, {\bf 107}, 219--263, 1999.

\bibitem{corm}
Cormen T.H., Leiserson C.H., Rivest R.L., {\em Introduction to algorithms}, MIT Press, Cambridge, 1990. 

\bibitem{Dicketal2008}
Dick U., Haider P., Scheffer T., Learning from Incomplete Data with Infinite Imputations, In: {\em Proceedings of the 25th International Conference on Machine Learning}, Helsinki, Finland, 2008.

\bibitem{dp}
Dubois D., Prade H., Fuzzy sets and probability: misunderstandings, bridges and gaps, In: {\em Second IEEE International Conference on Fuzzy Systems}, 1993.

\bibitem{green}
Green M.R., Pre-mRNA splicing. {\em Annu. Rev. Genet.}, {\bf 20}, 671--708, 1986.

\bibitem{haus}
Haussler D., Quantifying inductive bias: AI learning algorithms and Valiant's learning framework, {\em Artif. Intell.}, {\bf 36}, 177--222, 1988.

\bibitem{Ibrahim1990}
Ibrahim J., Incomplete data in generalized linear models. Journal of the American Statistical
Association, 85:765–769, 1990.

\bibitem{john}
Johnson D.S., Approximation algorithms for combinatorial problems, {\em J. Comput. Syst. Sci.}, {\bf 9}, 256--278, 1974.

\bibitem{kear}
Kearns M., Li M., Pitt L., Valiant L., On the learnability of Boolean formulae. In: {\em Proceedings of tghe 9th Annual ACM Symposium on Theory of Computing}, 285--295, 1987.

\bibitem{manga}
Mangasarian O.L., Setiono R., Wolberg W.H., Pattern recognition via linear programming: Theory and applications to medical diagnosis, In: T.E. Coleman, Y. Li (Eds.), {\em Large-Scale Numerical Optimization}, SIAM, Philadelphia, PA, 22--30, 1990.

\bibitem{mic}
Michalski R.S., A theory and methodology of inductive learning, {\em Artif. Intell.}, {\bf 20}, 111--116, 1983.

\bibitem{mit}
Mitchell T.M., Generalization as search, {\em Artif. Intell.}, {\bf 18}, 203--226, 1982.

\bibitem{pitt}
Pitt L., Valiant L.G., Computational limitations on learning from examples, {\em J. ACM}, {\bf 35}, 965--984, 1988.

\bibitem{ramp1}
Rampone S., Recognition of spline-junctions on DNA sequences by BRAIN learning algorithm, {\em Bioinformatics Journal}, {\bf 14}(8), 676--684, 1998.

\bibitem{ramp2}
Rampone S., An Error Tolerant Software Equipment For Human DNA Characterization, {\em IEEE Transactions on Nuclear Science}, {\bf 51}(5), 2018--2026, 2004.

\bibitem{rubin1987}
Rubin D. B., Multiple Imputation for Nonresponse in Surveys. John Wiley \& Sons, Inc., 1987.

\bibitem{val}
Valiant L.G., A theory of the learnable, {\em Comm. ACM}, {\bf 27}, 1134--1142, 1984.

\bibitem{Wangetal2006}
Wang X., Li A.,  Jiang Z., and Feng H., Missing value estimation for DNA microarray gene expression
data by support vector regression imputation and orthogonal coding scheme. BMC
Bioinformatics, 7:32, 2006.

\bibitem{wangetal2010}
Wang C., Liao X., Carin L., Dunson D. B., Classification with Incomplete Data Using Dirichlet Process Priors ; JMLR 11(Dec):3269-3311, 2010.

\bibitem{WilliamsandCarin2005} Williams D., and Carin L., Analytical kernel matrix completion with incomplete multi-view data. In
Proceedings of the International Conference on Machine Learning (ICML) Workshop on Learning
with Multiple Views, pages 80–86, 2005.

\bibitem{zad}
Zadeh L.A., Fuzzy sets, {\em Information and Control}, {\bf 8}/3, 338--353, 1965.

\bibitem{data}
Asuncion A., Newman D.J., UCI Machine Learning Repository, Irvine, CA.\\
http://www.ics.uci.edu/~mlearn/MLRepository.html\\
University of California, School of Information and Computer Science, 2007.











\end{thebibliography}
\end{document}